# Current-driven dynamics and inhibition of the skyrmion Hall effect of ferrimagnetic skyrmions in GdFeCo films

**Authors:** Seonghoon Woo,[1†*] Kyung Mee Song,[1,2†] Xichao Zhang,[3] Yan Zhou,[3] Motohiko Ezawa,[4] Xiaoxi Liu,[5] S. Finizio,[6] J. Raabe,[6] Nyun Jong Lee,[7] Sang-Il Kim,[7] Seung-Young Park,[7] Younghak Kim,[8] Jae-Young Kim,[8] Dongjoon Lee,[1,9] OukJae Lee,[1] Jun Woo Choi,[1,10] Byoung-Chul Min,[1,10] Hyun Cheol Koo,[1,9] and Joonyeon Chang[1,10]

**Affiliations:**

[1]Center for Spintronics, Korea Institute of Science and Technology, Seoul 02792, Korea

[2]Department of Physics, Sookmyung Women's University, Seoul 04130, Korea

[3]School of Science and Engineering, The Chinese University of Hong Kong, Shenzhen 518172, China

[4]Department of Applied Physics, University of Tokyo, Hongo 7-3-1, Tokyo 113-8656, Japan

[5]Department of Electrical and Computer Engineering, Shinshu University, Wakasato 4-17-1, Nagano 380-8553, Japan

[6]Swiss Light Source, Paul Scherrer Institut, 5232 Villigen, Switzerland

[7]Spin Engineering Physics Team, Division of Scientific Instrumentation, Korea Basic Science Institute, Daejeon, 305-806, Korea

[8]Pohang Accelerator Laboratory, Pohang University of Science and Technology, Pohang 37673, Korea

[9]KU-KIST Graduate School of Converging Science and Technology, Korea University, Seoul 02792, Korea

[10]Department of Nanomaterials Science and Engineering, Korea University of Science and Technology, Daejeon 34113, Korea

† These authors contributed equally to this work.

* Authors to whom correspondence should be addressed: shwoo_@kist.re.kr

**Magnetic skyrmions are swirling magnetic textures with novel characteristics suitable for future spintronic and topological applications. Recent studies confirmed the room-temperature stabilization of skyrmions in ultrathin ferromagnets. However, such ferromagnetic skyrmions show undesirable topological effect, the skyrmion Hall effect, which leads to their current-driven motion towards device edges, where skyrmions could easily be annihilated by topographic defects. Recent theoretical studies have predicted enhanced current-driven behaviour for antiferromagnetically exchange-coupled skyrmions. Here we present the stabilization of these skyrmions and their current-driven dynamics in ferrimagnetic GdFeCo films. By utilizing element-specific X-ray imaging, we find that the skyrmions in the Gd and FeCo sublayers are antiferromagnetically exchange-coupled. We further confirm that ferrimagnetic skyrmions can move at a velocity of ~50 m s$^{-1}$ with reduced skyrmion Hall angle, $\theta_{\text{SkHE}}$ ~20°. Our findings open the door to ferrimagnetic and antiferromagnetic skyrmionics while providing key experimental evidences of recent theoretical studies.**

**Introduction**

Magnetic skyrmions are non-trivial topological objects[1,2] that have been greatly highlighted recently, mainly due to their unique and fascinating topological characteristics suitable for future spintronic applications such as skyrmion-based racetrack memory.[3–5] Magnetic skyrmions can be stabilized in the presence of a strong Dzyaloshinskii-Moriya interaction (DMI),[6,7] which prefers non-collinear spin orientation between neighbouring magnetic moments. Recent investigations have revealed that, in structures where ultrathin ferromagnets are interfaced with large spin-orbit coupling materials, such as Ta/CoFeB/MgO,[8] Pt/Co/Ir,[9] Pt/Co/MgO,[10] Pt/Co/Ta and Pt/CoFeB/MgO,[11,12] the interfacial DMI can be strong enough to stabilize the chiral skyrmion structure even at room temperature. It has also been shown that skyrmions can move along magnetic tracks upon the injection of an electrical current,[8,11,13,14] which indicates that the skyrmions can be adopted in practical devices.[5] However, contrary to theoretical predictions, ferromagnetic skyrmions have shown relatively slow and pinning-dominated current-driven dynamic behaviours at room temperature.[8,11,13–15] More seriously, ferromagnetic skyrmions exhibit an inevitable topology-dependent effect, namely, the skyrmion Hall effect,[16–19] where the magnetic skyrmions do not move collinear to the current flow direction but acquire a transverse motion due to the appearance of a topological Magnus force acting upon the non-zero topological charge. The Magnus force is opposite for a ferromagnetic skyrmion whose topological charge ($Q$) is either $Q = 1$ or -1. To avoid this issue, theoretical studies have suggested that the skyrmion Hall effect can be suppressed by utilizing antiferromagnetically exchange-coupled skyrmions with $Q = 1$ and -1 in antiferromagnetic materials.[20–23] Along with the recent interest in antiferromagnets resulting from their intrinsic ultrafast dynamics and insensitivity to disturbing magnetic fields,[24] these advantages in skyrmion dynamics have generated an intense interest in antiferromagnetic skyrmion textures. However, since such an excitation behaves as if it were a simple bubble with $Q = 0$, it is quite difficult to experimentally verify if it is really a topological object.

In this work we investigate a similar object in ferrimagnet multilayers. Namely, we consider skyrmion excitations in one magnetic layer and in the magnetic other layer. Since they are antiferromagnetically coupled, their topological charges are opposite. Let us refer to such a pair of skyrmions as a ferrimagnetic skyrmion. The distinctive nature is the topological Magnus forces are not balanced exactly because the magnetization is different between two layers. We predict that ferrimagnetic skyrmions show a small but nonzero skyrmion Hall effect, which we may use as a verification of the topological excitation. We note a recent study reporting the stabilization of ferrimagnetic skyrmions in Fe/Gd multilayers,[25] in which Bloch-type skyrmions are stabilized by long-range dipolar interactions. However, current-driven dynamics of these skyrmions cannot be used for actual applications, mainly due to the random chirality of each individual skyrmion that results in a complex and random current-driven motion.[26,27] Thus, the actual observation of deterministic and efficient current-driven dynamics of chiral ferrimagnetic skyrmions remains elusive so far and needs to be pursued for the realization of skyrmion-based memory and logic computing devices.

**Results**

**X-ray microscopy observation of magnetic domains**

In our multilayer stack, the Pt heavy metal underlayer is used to induce a strong DMI that stabilizes chiral magnetic textures. Note that the [Pt(3 nm)/$Gd_{25}Fe_{65.6}Co_{9.4}$(5 nm)/MgO(1 nm)]$_n$ structure with a large repetition number, $n = 20$, is used in this study, due to: i) the sizable internal demagnetization field that effectively drives as-grown the magnetization state into the multi-domain state (see Supplementary Fig. 1 and Supplementary Note 1 for its hysteresis behaviours and the top left panel in Fig. 1a for its as-grown multi-domain state), ii) a large amount of magnetic material that resulted in the enhanced magnetic contrast in our X-ray transmission measurement, and iii) promising skyrmion characteristics observed in the structure, which will be presented throughout this Article. Using vibrating sample magnetometry (VSM) measurements, we estimate the magnetization compensation point, $T_M$, of our GdFeCo film to be > 450 K (see Supplementary Fig. 2 and Supplementary Note 2 for details). Hence, our magnetic devices remain at an uncompensated ferrimagnetic state throughout all the room-temperature X-ray measurements. Moreover, to correctly extract the magnetic parameters of our ferrimagnetic films, we have performed several experiments including ferromagnetic resonance (FMR), spin-torque FMR (ST-FMR), and X-ray magnetic circular dichroism (XMCD) spectroscopy measurements on a companion GdFeCo film grown on $SiO_x$/Si substrate (see Supplementary Figs. 3-6, Supplementary Note 3 and Methods for detailed measurement descriptions and acquired parameter values).

To reveal the nature of ferrimagnetic skyrmions in our system, we first performed element-specific scanning transmission X-ray microscopy (STXM) in the presence of an external perpendicular magnetic field, $B_z$. Figure 1a shows the STXM images of the domain structure in a patterned 2.5-μm-wide and 5-μm-long Pt/GdFeCo/MgO film with decreasing external magnetic field from $B_z = 0$ mT to $B_z = -130$ mT. The upper and lower panels show corresponding STXM images taken at the absorption edges of Fe ($L_3$-edge) and Gd ($M_5$-edge), respectively. In these STXM images, dark and bright contrasts correspond to upward (+$M_z$) and downward (-$M_z$) magnetization directions, respectively. At $B_z$ = 0 mT, a labyrinth stripe domain state with the average domain width of ~220 nm is achieved. Moreover, it is immediately obvious that STXM images at Fe- and Gd-absorption edges show opposite contrast, revealing their expected antiferromagnetic spin ordering within the GdFeCo alloy. Note that, since the measurements are conducted at room temperature, which is lower than the compensation point $T_M$, the magnetic moment of Gd aligns parallel with the external magnetic field while the moment of Fe aligns in an anti-parallel fashion.[28] As the magnetic field increases, fewer domains remain and magnetic configurations become less complex. Eventually, by reaching a magnetic field of $B_z$ = -130 mT, we observe multiple isolated skyrmions, and it is evident that the Gd and Fe magnetic moments are still antiferromagnetically exchange-coupled within these skyrmions, therefore confirming that we observed ferrimagnetic skyrmions. The high spatial resolution (~25 nm) of STXM allows us to measure the diameter of the observed skyrmions as discussed in Supplementary Figs. 7-8 and Supplementary Note 4, and we find that the average skyrmion diameter is roughly ~180 nm, which is as small as the skyrmions found in ferromagnetic multilayers with a large DMI value, 1.5 ~ 2 mJ m$^{-2}$, studied in Refs. [9,11,12]. We later confirm that our skyrmions exhibit a left-handed Néel-type chirality by observing their current-driven behaviours. Figure 1b schematically illustrates the orientation of the antiferromagnetically exchange-coupled internal magnetic moments within the observed ferrimagnetic skyrmion structure.

**Current-driven behaviours of ferrimagnetic skyrmions**

Having established that ferrimagnetic skyrmions can form at a finite external field in this material, we next study their current-induced dynamics in the magnetic track. Figure 2a shows a schematic drawing of our ferrimagnetic track and electric contacts patterned on a 100-nm-thick $Si_3N_4$ membrane for transmission X-ray measurements. The actual scanning electron microscopy (SEM) micrograph of our device is also included, and two indicated areas within the image, (i) and (ii), are used to analyse current-driven skyrmion behaviours shown in Fig. 2b and 2c, respectively. In Fig. 2b and 2c, each STXM image, taken at Fe-edge, was acquired after injecting a single current pulse, with the various pulse amplitudes of between $4.90\times10^{10}$ A m$^{-2}$ ≤ $|j_a|$ ≤ $3.55\times10^{11}$ A m$^{-2}$ and the pulse duration of 5 ns (see Supplementary Fig. 9 and Supplementary Note 5 for details on the electronic connections and the actual pulse shape). Note that each skyrmion is colour-circled, and a single colour is used for the same skyrmion throughout each sequence. The pulse polarity, defined by the electric current flow direction, is indicated in the figure as red and blue pulse-shaped arrows. It is also noteworthy that, while skyrmion core points +z and –z directions in STXM images in Fig. 2b and Fig. 2c, respectively, the effective core magnetization points -z and +z directions in Fig. 2b and Fig. 2c, respectively, because Gd moments are dominant in our material at room temperature as discussed above.

Fig. 2b first shows a sequence of STXM images of skyrmions stabilized by a magnetic field of $B_z$ = 145 mT near the left Au electrode on a magnetic track. In our field of view, there are initially two skyrmions, and as we inject leftward current pulses, additional two skyrmions appear while all of the skyrmions show homogeneous propagations. Moreover, it is noticeable that, when a train of the skyrmions propagates along the track, their alignment/trajectory shows a finite angle with respect to the current flow direction, which is the hallmark of the skyrmion Hall effect. We then reversed the magnetic field to $B_z$ = -145 mT and investigated another region near the right Au electrode on the same track, as shown in Fig. 2c. While no skyrmion is observed in the field of view of the first image, up to four skyrmions appear with pulse injections, and all of them show pinning-free homogeneous displacements, as was observed for the other polarity of skyrmions in Fig. 2b. Note that a train of these bright skyrmions also shows transverse velocity component, and surprisingly, the sign of slope is opposite to the case of dark skyrmions. This is due to the opposite topological polarity of skyrmions for two cases, which experience the opposite sign of the topological Magnus force that consequently provides opposite transverse propagation directions. This observation of finite skyrmion Hall effect and the symmetry of skyrmion Hall angle in our ferrimagnetic material agrees well with the cases of ferromagnetic skyrmions in Ta/CoFeB/TaO$_x$[17] and Pt/CoFeB/MgO.[18] Moreover, in the last image of each sequence, we observe the repulsion between skyrmions and the edges of the sample, which locates right underneath the field of views as shown in Fig. 2a. The repulsion occurs due to the DMI boundary condition,[29] which results in the skyrmion motion i) back toward the sample center (blue-circled skyrmion in Fig. 2b) or ii) straight along the sample edge (yellow-circled skyrmion in Fig. 2b and blue-circled skyrmion in Fig. 2c), which also agrees with previously observed ferromagnetic skyrmion motion along the edge in Ref. [17].

With these observations shown in Fig. 2b and 2c, three important qualitative conclusions on the skyrmion physics within ferrimagnetic material can be drawn. First and most importantly, our investigation reveals that ferrimagnetic skyrmions can also be displaced by electric currents at room temperature just as the ferromagnetic skyrmions.[8,11–

[14,17,18] Moreover, we show that the skyrmion propagation direction is along the current flow direction (against the electron flow) for both +$M_z$-core and -$M_z$-core skyrmions, and this same directionality agrees well with the spin Hall current-driven motion of homochiral left-handed Néel-type hedgehog skyrmions stabilized by interfacial DMI in Pt/ferromagnet thin films.[11–14,18] This implies that the interfacial DMI at the Pt/GdFeCo interface plays a crucial role in stabilizing skyrmions and also driving them on the track in our ferrimagnetic structure. Furthermore, the skyrmion pinning, which was often observed in many of ferromagnetic systems,[11,13,14,17] is significantly reduced in our ferrimagnetic material. This low pinning may originate from the amorphous nature of GdFeCo because the absence of grain boundaries leads to lower skyrmion pinning, as was observed in amorphous CoFeB ferromagnetic films.[11] Overall, it is noteworthy that our observation serves as the first experimental observation of current-driven excitation of nanoscale magnetic skyrmions in ferrimagnets.

The pulse amplitude-dependent skyrmion velocity and its skyrmion Hall angle are plotted in Fig. 2d and 2e, respectively. To correctly calculate the distance and angle between two images, we have performed image-displacement correction using the edge between our magnetic track and Au electrode. It is first noticeable that skyrmion velocity increases linearly with pulse amplitudes, and the maximum velocity approaches ~50 m s$^{-1}$ at $|j_a|$ = 3.55×10$^{11}$ A m$^{-2}$, which is comparable to the current state-of-the-art skyrmions observed in a few ferromagnetic heterostructures.[11,14,18] Moreover, as shown in Fig. 2e, we observe a very small skyrmion Hall angle, $|\theta_{SkHE}|$ up to ~20°, which is far lower than the skyrmion Hall angles, $|\theta_{SkHE}|$ > 30°, observed for ferromagnetic skyrmions in Ta/CoFeB/MgO and Pt/CoFeB/MgO structures.[17,18] Antiferromagnetic coupling between two sublayers and the corresponding largely-reduced net magnetization within GdFeCo films have led to the effective inhibition of the skyrmion Hall effect. It is noteworthy that relatively large skyrmion Hall effect in conventional ferromagnets may lead skyrmions toward device edges, where they could easily be annihilated by topographic defects.[20,30] Moreover, the skyrmion Hall effect-driven strong transverse motion may pose a limitation for the maximum skyrmion speed due to the finite edge repulse. Therefore, we believe that our ferrimagnetic multilayers can serve as an important magnetic material for such future skyrmionic devices that could replace conventional ferromagnets with enhanced reliability and mobility. Note that skyrmion Hall angle increases monotonically at low current densities, $|j_a|$ < ~2×10$^{11}$ A m$^{-2}$, and saturates at high current densities, because the skyrmion dynamics is dominated by pinning sites at low driving forces, which is similar to the creep motion of ferromagnetic skyrmions in low-current-density regime caused by the pinning potential.[15,17] The remnant finite skyrmion Hall angle, $\theta_{SkHE}$ ~ 20°, results from the uncompensated magnetic moments between Gd and FeCo sublayers at 300 K < $T_M$, where $M_{S\_Gd} \neq M_{S\_FeCo}$. (see Methods). Therefore, by adjusting material compositions, it will be possible to further reduce the effective skyrmion Hall angle in ferrimagnetic GdFeCo films.

**Micromagnetic simulation on ferrimagnetic skyrmion dynamics**

For more comparison, we simulated the current-driven dynamics of a ferrimagnetic skyrmion (see Methods for more modelling details) in a checkerboard-like two-sublayer spin system based on the G-type antiferromagnetic structure with simple square lattices,[20,30,31] where the two sublayers, corresponding to Gd and FeCo, are coupled in a ferrimagnetic manner with a net saturation magnetization, while each sublayer is ferromagnetically ordered. We have also examined simulations using the two-sublattice model with classic $J_1$-$J_2$-$J_2$' Heisenberg exchange interactions as shown in Supplementary Figs. 10-12 and Supplementary Note 6. While we find that the intra-sublattice exchange interactions indeed affect on the ferrimagnetic skyrmion size and dynamics, however, the influence of these effects on the overall dynamics, especially on the skyrmion Hall effect, turns out not to be significant in both qualitative and quantitative results.

Simulations were performed with both models: with and without pinning defects, using experimentally measured materials parameters given in Methods. Moreover, in simulations, we considered the error range of damping coefficient measurement shown in Supplementary Fig. 3 and Supplementary Note 3, because the ferrimagnetic skyrmion dynamics can be strongly influenced by small changes in damping coefficient (see Supplementary Figs. 13-14 and Supplementary Note 7 for details). Note that the error ranges of simulations are shown as shaded areas in Fig. 2d and 2e. The simulated skyrmion velocity as a function of the current density is first shown in Fig. 2d. Simulation results show qualitative and quantitative agreement with experimental observations, revealing that the ferrimagnetic skyrmion velocity is linearly proportional to the driving current density. We also calculated the skyrmion Hall angle as a function of the current density as shown in Fig. 2e. It is first noticeable that the skyrmion Hall angle in the model without pinning defects is independent of the current density, while the skyrmion Hall angle in the model with certain pinning defects increases with increasing current density and approaches a constant value calculated with the pinning-absent model. This linear increase is qualitatively consistent with our experimental observations and also with previous report[17], indicating the existence of certain pinning effects due to impurities or defects in the real material. Moreover, the calculated skyrmion Hall angle considering the damping errors shows a fair quantitative agreement with experimental observation.

Although the averaged skyrmion Hall angle for simulation is still slightly larger than experiments, we speculate that the small difference may originate from the larger effective damping associated with skyrmion dynamics. Weindler *et al.* recently reported that local FMR ($\alpha$ = 0.0072) and domain wall dynamics measurements ($\alpha$ = 0.023) yield very different damping parameters for the same material, Permalloy, and magnetic texture-induced nonlocal damping may be responsible for the increase in effective damping.[32] Gerrits *et al.* also reported that, unlike small-angle magnetization dynamics such as conventional FMR, large-angle magnetization dynamic could induce an increase in the apparent damping,[33] and we believe this scenario could also be used to explain our case, where skyrmion motion involves the large-angle magnetization dynamics. Moreover, because a patterned 2.5-μm-wide and 5-μm-long nanowire structure was used for skyrmion study while continuous films were employed in material parameter analysis, roughness-induced extrinsic damping enhancement could be another source of damping increase.[34] Nevertheless, by considering above possible scenarios, our quantitative results on the suppression of the skyrmion Hall effect can be reasonably understood.

**Discussion**

We have so far observed that the current-driven behaviours of ferrimagnetic skyrmions are indeed attractive compared with their ferromagnetic counterparts. However, it may also be possible that the small value of skyrmion Hall angle in a ferrimagnetic material is only from the small saturation magnetization in a ferrimagnet and not from its antiferromagnetic characteristic. Thus, for a fair comparison between ferrimagnetic and ferromagnetic skyrmions, we have performed simulations on the current-driven dynamics of skyrmions with the same low (net) saturation magnetization, as shown in Fig. 3. It should first be pointed out that, by using the same material parameters, the sizes of the ferrimagnetic and ferromagnetic skyrmions are measured to be identical at certain out-of-plane magnetic fields (Fig. 3a, inset). Figure 3a shows the velocities of the current-driven ferrimagnetic and ferromagnetic skyrmions as a function of the driving current density. It can be seen that both the ferrimagnetic and ferromagnetic skyrmion velocities increase with increasing driving current density, where the mobility of ferrimagnetic skyrmion is slightly larger than that of ferromagnetic skyrmion. More significantly, as shown in Fig. 3b, the skyrmion Hall angle of ferrimagnetic skyrmion is significantly smaller than that of the ferromagnetic skyrmion by a factor of 2, even the net saturation magnetizations are the same for both skyrmions. Note that skyrmion Hall angles are independent of the current densities for both cases, as the pinning effect is not considered in this comparison. To compare the effect of pinning on ferrimagnetic and ferromagnetic skyrmions more carefully, we have simulated the current-driven dynamics of both ferrimagnetic and ferromagnetic skyrmions in the same disorder model, which is shown in Supplementary Fig. 15 and Supplementary Note 8. It is found that the given pinning effect on the current-driven skyrmion dynamics is not significant for both ferrimagnetic and ferromagnetic cases, especially at a large driving current density (e.g. $5\times10^{11}$ A m$^{-2}$), however, at a small driving current density (e.g. $1\times10^{11}$ A m$^{-2}$), the ferrimagnetic skyrmion motion is more influenced by the pinning effect. The reason could be that the ferromagnetic skyrmion experiences a stronger Magnus force, which helps it in overcoming obstacles.[35,36] Nevertheless, larger skyrmion velocity and smaller skyrmion Hall angle for ferrimagnetic case were maintained over the whole range of examined current densities. Overall, simulation results indicate that, even when the ferrimagnetic and ferromagnetic skyrmions have the same low (net) saturation magnetization, the current-driven dynamics of the ferrimagnetic skyrmion is much more reliable for the transport in narrow channels as information carriers.

In conclusion, we have observed and studied the stabilization and current-driven dynamics of antiferromagnetically exchange-coupled skyrmions in ferrimagnetic GdFeCo films. By utilizing the element-specific X-ray imaging, we have identified that the ferrimagnetic skyrmion in the GdFeCo films consists of two antiferromagnetically exchange-coupled skyrmions in the Gd and FeCo sublayers. We further confirm that current-driven ferrimagnetic skyrmions can move at a velocity of ~50 m s$^{-1}$ with reduced skyrmion Hall angle, $|\theta_{SkHE}|$ ~20°. With micromagnetic simulations, we reveal that ferrimagnetic skyrmions are much more attractive than their ferromagnetic counterparts in many technological-relevant aspects, such as larger skyrmion mobility and strongly suppressed skyrmion Hall effect, mainly due to their antiferromagnetic nature. Our findings reveal the promising dynamic properties of ferrimagnetic skyrmions, and highlight the possibility to build more reliable skyrmionic devices using ferrimagnetic and antiferromagnetic materials.

## Methods

### Sample preparation and experimental method

The [Pt(3 nm)/Gd$_{25}$Fe$_{65.6}$Co$_{9.4}$(5 nm)/MgO(1 nm)]$_{20}$ films were grown by DC magnetron sputtering at room temperature under 1 mTorr Ar for Pt and GdFeCo and 4 mTorr Ar for MgO at a base pressure of roughly ~2×10$^{-8}$ Torr. Samples were grown on a 100-nm-thick SiN substrate and then patterned using electron beam lithography and lift-off technique. Nominally sample films were grown on SiOx/Si substrate for vibrating-sample magnetometry (VSM), ferromagnetic resonance (FMR), spin-torque FMR (ST-FMR), asymmetric bubble expansion measurements. The series of measurements yielded material constants: anisotropic field $\mu_0 H_k$ = 0.15 T, net saturation magnetization $M_S$ = 2×10$^5$ A m$^{-1}$, uniaxial anisotropy, $K_u$ = 4.01×10$^4$ J m$^{-3}$, damping coefficient, $\alpha$ = 0.205±0.035, spin Hall angle, $\theta_{SH}$ = 0.055, magnetization ccompensation ratio, $n = M_S^{FeCo}/M_S^{Gd}$ = 0.78 and DMI magnitude, $D$ = -0.98 mJ m$^{-2}$ (see Supplementary Note 3 for details).

All microscopy images were acquired using the STXM installed at the PolLux (X07DA) beamline of the Swiss Light Source (SLS) at the Paul Scherrer Institute in Villigen, Switzerland. The device used for the experiments was 2.5-μm-wide and 5-μm-long, which yielded an electrical resistance of ~57 Ohms measured in 2-point. This resistance reduced the impedance mismatch of the device, and allowed an almost complete transmission of 5-ns-long short electrical pulses across the device. This was verified by simultaneously measuring the injected (through a -20 dB pickoff T) and transmitted samples with a real-time oscilloscope. Pulse current densities above ~4×10$^{11}$ A m$^{-2}$ led to a damage of the Au contact, which eventually limited the maximum current applied in Fig. 2. Skyrmion velocities were determined using the total displacements, measured by acquiring XMCD-STXM images before and after the injection of the pulses, and the integrated pulse time. Three to ten displacements were recorded for each pulse amplitude, and the average value and standard deviations of the individual velocity measurements are plotted in Fig. 2d. Current densities were calculated by dividing the injected current with stripe width and effective total thickness of Pt and GdFeCo.

### Simulation method

The spin dynamics simulation is carried out by using the Object Oriented MicroMagnetic Framework (OOMMF) with the home-made extension modules for the periodic boundary condition.[37] The model is treated as a checkerboard-like two-sublattice spin system based on the G-type antiferromagnetic structure with simple square lattices, where the two sublattices are coupled in a ferrimagnetic manner with a net spontaneous magnetization, while each sublattice is ferromagnetically ordered. The Hamiltonian is based on the classical Heisenberg model, given as

$$\mathcal{H} = -J_{ij}\sum_{<i,j>}\mathbf{S}_i \cdot \mathbf{S}_j - J_{ij}^{A}\sum_{\ll i,j \gg_A}\mathbf{S}_i \cdot \mathbf{S}_j - J_{ij}^{B}\sum_{\ll i,j \gg_B}\mathbf{S}_i \cdot \mathbf{S}_j + D\sum_{<i,j>}(\mathbf{u}_{ij} \times \hat{z}) \cdot (\mathbf{S}_i \times \mathbf{S}_j)$$
$$-K\sum_i (S_i^z)^2 - \sum_i \mu_i \mathbf{S}_i \cdot \mathbf{H} + H_{\text{DDI}} \quad (1)$$

where $\mathbf{S}_i$ represents the local spin vector reduced as $\mathbf{S}_i = \mathbf{M}_i/M_S^i$ at the site $i$, and $\mathbf{S}_j$ represents the local spin vector reduced as $\mathbf{S}_j = \mathbf{M}_j/M_S^j$ at the site $j$. $\mathbf{M}_i$ and $\mathbf{M}_j$ are the magnetization at the site $i$ and $j$, respectively. $M_S^i$ denotes the saturation magnetization of the sublattice $i$, while the saturation magnetization of sublattice $j$ is defined as $M_S^j = nM_S^i$ with the compensation ratio $n$. $<i, j>$ runs over all the nearest-neighbor sites in the two-sublattice spin system. $\ll i, j \gg_A$ and $\ll i, j \gg_B$ run over all the nearest-neighbor sites in the sublattice A and sublattice B, respectively (see Supplementary Fig. 10). $J_{ij}$ is the exchange coupling energy constant between the two adjacent spin vector $\mathbf{S}_i$ and $\mathbf{S}_j$, which has a negative value ($J_{ij} < 0$) representing the antiferromagnetic spin ordering of the two sublattices. $J_{ij}^A$ and $J_{ij}^B$ are the exchange coupling energy constants for the sublattice A and sublattice B, respectively, which are positive numbers ($J_{ij}^A > 0$, $J_{ij}^B > 0$) representing the ferromagnetic intra-sublattice coupling. $D$ is the interface-induced DMI constant, $\mathbf{u}_{ij}$ is the unit vector between spins $\mathbf{S}_i$ and $\mathbf{S}_j$, and $\hat{z}$ is the interface normal, oriented from the heavy-metal layer to the ferrimagnetic layer. $K$ is the perpendicular magnetic anisotropy (PMA) constant, $\mathbf{H}$ is the applied magnetic field, and $H_{\text{DDI}}$ stands for the dipole-dipole interaction, i.e., the demagnetization effect.

The time-dependent dynamics of the spin system is controlled by the Landau-Lifshitz-Gilbert (LLG) equation augmented with the damping-like spin Hall torque, which is expressed as

$$\frac{d\mathbf{S}_i}{dt} = -\gamma_0 \mathbf{S}_i \times \mathbf{H}_{\text{eff}} + \alpha\left(\mathbf{S}_i \times \frac{d\mathbf{S}_i}{dt}\right) + \tau[\mathbf{S}_i \times (\hat{p} \times \mathbf{S}_i)] \quad (2)$$

where $\mathbf{H}_{\text{eff}} = -(1/\mu_0 M_S^i) \cdot (\delta\mathcal{H}/\delta\mathbf{S}_i)$ is the effective field on a lattice site, $\gamma_0$ is the Gilbert gyromagnetic ratio, and $\alpha$ is the phenomenological damping coefficient. The coefficient for the spin Hall torque is given as $\tau = (\gamma_0 \hbar j \theta_{\text{SH}})/(2\mu_0 e M_S^i b)$, where $j$ is the applied charge current density, $\theta_{\text{SH}}$ is the spin Hall angle, and $b$ is the thickness of the ferrimagnetic layer. $\hat{p} = \mathbf{j} \times \hat{z}$ denotes the spin polarization direction.

For the simulation on the multilayer structure, we employed an effective medium approach[11] with the lattice constant of 5 Å, which improves the computational speed by converting the multilayer into a two-dimensional effective model with reduced parameters. The intrinsic magnetic parameters used in the simulation are measured from our experimental samples as well as adopted from Refs. [11,18,38]: the damping coefficient $\alpha = 0.205 \pm 0.035$, the inter-sublattice exchange stiffness $A_{\text{Gd-Fe}}$ = -10 pJ m$^{-1}$, $A_{\text{Gd-Gd}}$ = 5 pJ m$^{-1}$, $A_{\text{Fe-Fe}}$ = 5 pJ m$^{-1}$, the spin Hall angle $\theta_{\text{SH}} = 0.055$, the DMI constant $D$ = -0.96 mJ m$^{-2}$, the PMA constant $K_u$ = 4.01×10$^4$ J m$^{-3}$, and the net saturation magnetization $M_S = M_S^{\text{Gd}} - M_S^{\text{Fe}} = 200$ kA m$^{-1}$. The compensation ratio is measured as $n = M_S^{\text{Fe}}/M_S^{\text{Gd}} = 0.78$. In the experimental multilayer system, the thickness of one ferrimagnetic layer is $t_m$ = 5 nm, the thickness of one repetition is $t_r$ = 9 nm, and the number of repetitions is $n_{\text{rep}}$ = 20. For the simulation on the model with pinning defects, the defects with the size of 10 Å × 10 Å and a higher PMA ($K_p = 5K_u$) are randomly distributed in the whole ferrimagnetic layer. The density of the defects in the whole model equals 5 %.

**Data availability**

Data supporting the findings of this study are available within the article and its Supplementary Information files and from the corresponding author upon request. The micromagnetic simulator OOMMF used in this work is publicly accessible at http://math.nist.gov/oommf.

**Acknowledgements**

This work was primarily supported by Samsung Research Funding Center of Samsung Electronics under Project Number SRFC-MA1602-01. Part of this work was performed at the PolLux (X07DA) beamline of the Swiss Light Source. S.W. and J.W.C. acknowledge the support from KIST Institutional Program. K.M.S acknowledges the support from the Sookmyung Women's University BK21 Plus Scholarship. X.Z. was supported by JSPS RONPAKU (Dissertation Ph.D.) Program. Y.Z. acknowledges the support by the President's Fund of CUHKSZ, the National Natural Science Foundation of China (Grant No. 11574137), and Shenzhen Fundamental Research Fund (Grant Nos. JCYJ20160331164412545 and JCYJ20170410171958839). M.E. acknowledges the support by the Grants-in-Aid for Scientific Research from JSPS KAKENHI (Grant Nos. JP17K05490, 25400317 and JP15H05854), and also the support by CREST, JST (Grant No. JPMJCR16F1). S.F acknowledges the support by the EU Horizon 2020 MAGicSky project (Grant No. 665095). K.M.S. and J.W.C. acknowledge the travel fund supported by the National Research Foundation of Korea (NRF) funded by the MSIP (2016K1A3A7A09005418). S.-Y.P. and B.-C.M. acknowledge the support from the National Research Council of Science & Technology (NST) grant (No. CAP-16-01-KIST) by the Korea government (MSIP). S.-Y.P. also acknowledges the support from KBSI Grant (D37614). S.W. also acknowledges S. Emori for his helpful comments on the manuscript.

**Author Contributions**

S.W. designed and initiated the study. K.M.S. optimized structure, fabricated devices and performed the film characterization. S.W., K.M.S., S.F. and J.R. performed X-ray imaging experiments using STXM at Swiss Light Source in Villigen, Switzerland. X.Z. and Y.Z. performed the numerical simulations. M.E. carried out the theoretical analysis. During the revision of this article, N.J.L., S.-I.K. and S.-Y.P. performed ferromagnetic resonance (FMR), D.L. and O.L. performed spin-torque FMR, and S.W., K.M.S., Y.H.K., J.-Y.K. and J.W.C. performed XMCD spectroscopy at 2A beamline at Pohang Accelerator Laboratory in Pohang, Korea. S.W., X.Z. and M.E. drafted the manuscript and revised it with assistance from X.L., D.L., O.L., S.-Y.P., J.W.C., B.-C.M., H.C.K. and J.C.. All authors commented on the manuscript.

**Competing Financial Interest**

The authors declare no competing financial interests.

**Author Information**

S.W. and K.M.S. contributed equally to this work. Correspondence and requests for materials should be addressed to S.W. (shwoo_@kist.re.kr).

**Figure Legends**

**Figure 1. Scanning transmission X-ray microscopy (STXM) imaging of domain structure upon magnetic field application. a,** STXM images acquired by sweeping the external perpendicular magnetic field from $B_z = 0$ mT to $B_z = -130$ mT. Dark and bright contrasts correspond to magnetization oriented up (along +$z$) and down (along -$z$), respectively. Upper panel and lower panel show corresponding images acquired at the $L_3$ and $M_5$ absorption edges of Fe and Gd, respectively. Note that, due to longer penetration depth associated with the higher energy used for Gd, ~1189 eV, compared with that of Fe, ~709 eV, magnetic contrast under Au electrodes is visible for Gd magnetic moment imaging. **b,** Schematic of antiferromagnetically exchange-coupled ferrimagnetic skyrmion on a magnetic track as observed in our GdFeCo films as indicated in the red dashed-square boxes in the last image of **a.** Scale bar, 1 μm.

**Figure 2. Current-driven behaviour of ferrimagnetic skyrmions and their velocity and skyrmion Hall effect. a,** Schematic of scanning transmission X-ray microscopy (STXM) geometry, and a scanning electron microscopy (SEM) image of the actual device used for experiments. Scale bar, 2 μm. Sequential STXM images taken at Fe-edge showing the responses of multiple skyrmions after injecting unipolar current pulses along the track at **b,** $B_z = 145$ mT and **c,** $B_z = -145$ mT, respectively. With a fixed pulse-length of single pulse, 5 ns, the pulse amplitude is changed between $4.90 \times 10^{10}$ A m$^{-2}$ ≤ |$j_a$| ≤ $3.55 \times 10^{11}$ A m$^{-2}$. Pulse polarities are indicated as red- and blue-coloured arrows inside each image. Within STXM images, the same skyrmion is indicated with the same colour. Scale bars, 500 nm. **d,** Experimental and simulated average skyrmion velocity of Pt/GdFeCo/MgO versus current density. **e,** Experimental and simulated average skyrmion Hall angle of Pt/GdFeCo/MgO versus current density. In **d-e,** The shaded areas in plots represent the simulation results considering the damping coefficient error ranges of $\alpha = 0.205 \pm 0.035$, which was measured experimentally as described in Supplementary Fig. 3 and Supplementary Note 3. Note that pulse current densities above ~$4 \times 10^{11}$ A m$^{-2}$ led the damage of the Au contact, which eventually limited the maximum applicable current to our sample. Error bars denote the standard deviation of multiple measurements.

**Figure 3. Comparison between the current-driven dynamics of ferrimagnetic and ferromagnetic skyrmions of the same net saturation magnetization.** **(a)** The ferrimagnetic and ferromagnetic skyrmion velocities as a function of the driving current density. Inset shows the close-up top view of the ferrimagnetic skyrmion in a square film with periodic boundary conditions in both $x$ and $y$ directions. The lattice constant is set as 5 Å. **(b)** The ferrimagnetic and ferromagnetic skyrmion Hall angle as a function of the driving current density. The net saturation magnetization was set to be $M_S = 2\times10^5$ A m$^{-1}$ for both cases. Scale bar, 5 nm.

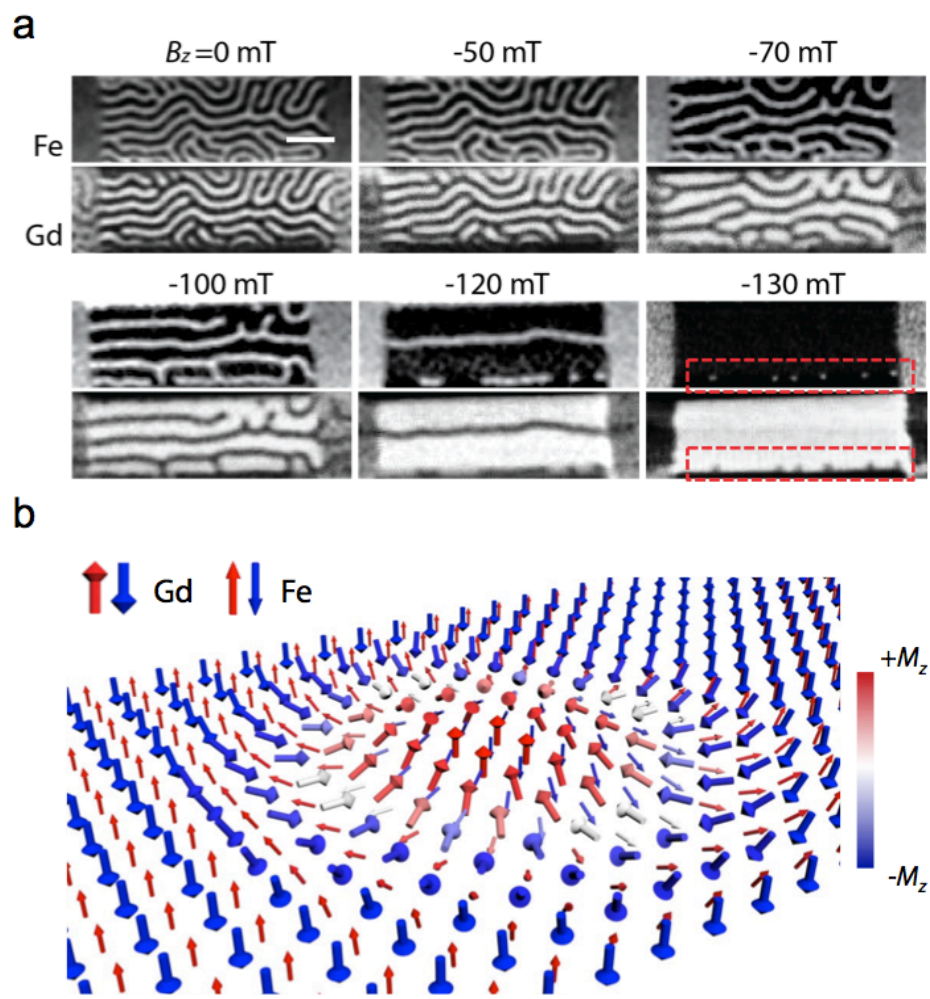

Figure 1

Seonghoon Woo *et al.*

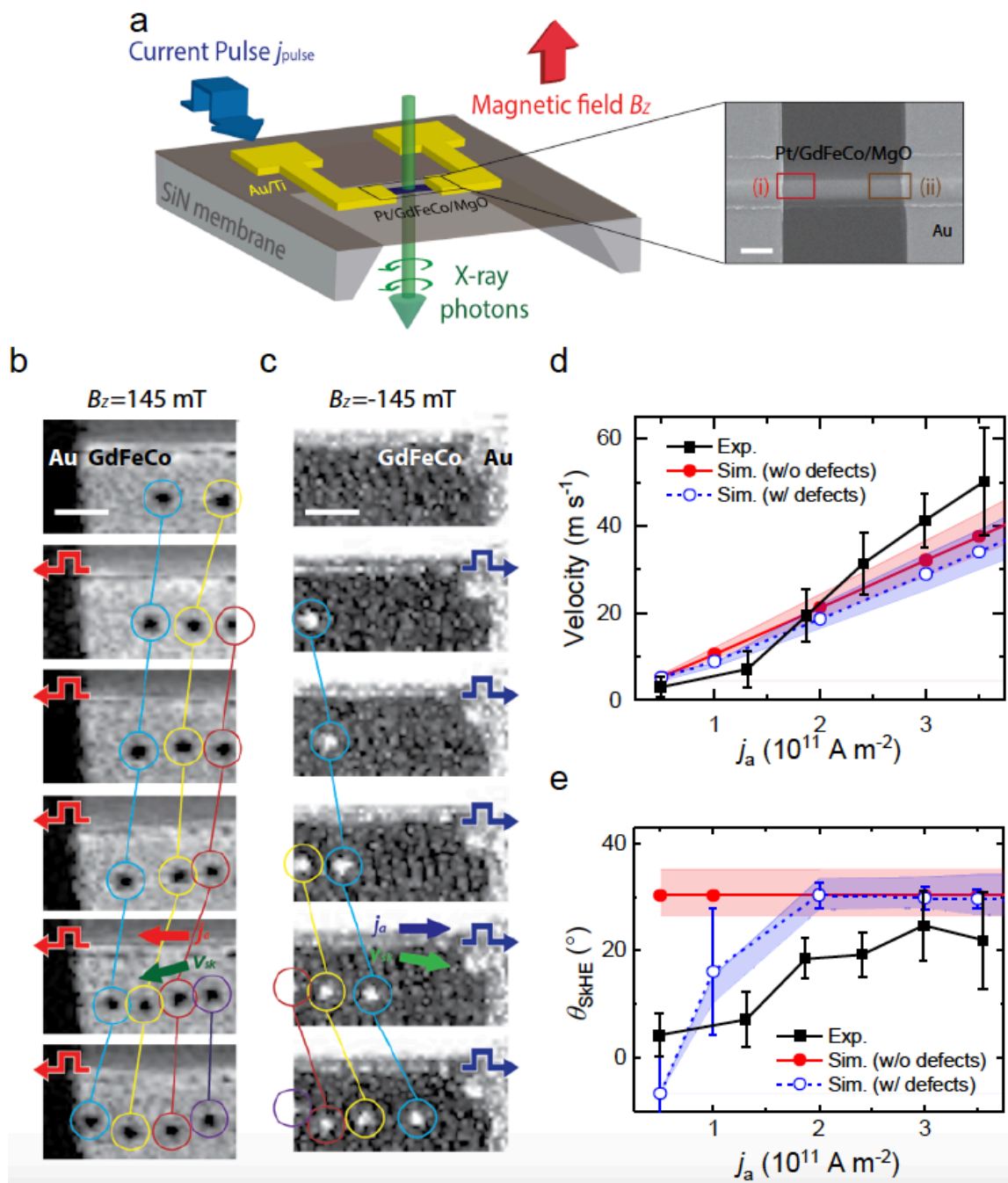

Figure 2

Seonghoon Woo *et al.*

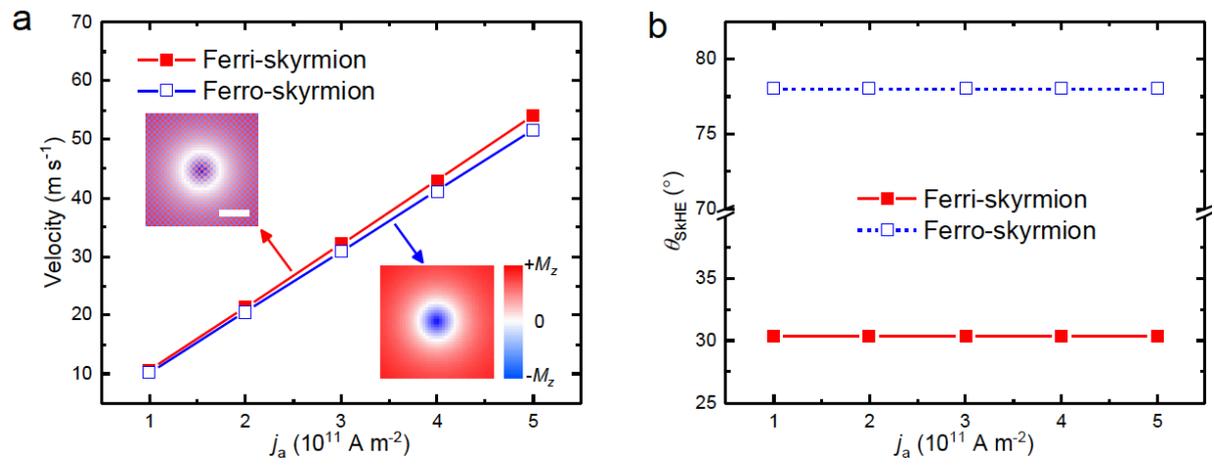

Figure 3

Seonghoon Woo *et al.*